\documentclass[preprint,12pt]{elsarticle}
\usepackage{hyperref}
\usepackage{amsmath}
\usepackage{amssymb}
\usepackage{cancel}
\usepackage{xspace}
\newcommand{\NASixtyOne}{NA61\slash SHINE\xspace}
\newcommand{\NAFortyNine}{NA49\xspace}

\makeatletter
\newcommand{\@reaction}[2]{\mbox{\ensuremath{#1+#2}}\xspace}
\newcommand{\NewReaction}[3]{
  \newcommand{#1}{\@reaction{#2}{#3}}
}
\makeatother

\newcommand{\NewMathSymbol}[2]{
  \newcommand{#1}{\ensuremath{#2}\xspace}
}

\newcommand{\NewIon}[2]{
  \NewMathSymbol{#1}{\text{#2}}
}

\NewIon{\Ar}{Ar}
\NewIon{\Be}{Be}
\NewIon{\Sc}{Sc}
\NewIon{\Pb}{Pb}
\NewIon{\Carbon}{C}
\NewIon{\Silicon}{Si}
\NewIon{\AnyIon}{A}

\NewReaction{\PbPb}{\Pb}{\Pb}
\NewReaction{\CC}{\Carbon}{\Carbon}
\NewReaction{\SiSi}{\Silicon}{\Silicon}
\NewReaction{\ArSc}{\Ar}{\Sc}
\NewReaction{\BeBe}{\Be}{\Be}
\NewReaction{\AnyReaction}{\AnyIon}{\AnyIon}

\newcommand{\crange}[2]{\mbox{\ensuremath{#1-#2\%}}\xspace}
\newcommand{\cms}{\mbox{\ensuremath{\sqrt{s_{NN}}}}\xspace}

\journal{Nuclear Physics A}

\begin{document}

\begin{frontmatter}

\title{\mbox{Decoding the QCD critical behaviour in \AnyReaction collisions}\tnoteref{dedicatory}}
\tnotetext[dedicatory]{Dedicated to the memory of N.~G.~Antoniou (1939-2020)}

\author[uoa]{N.~G.~Antoniou\corref{dec}}
\ead{nantonio@phys.uoa.gr}

\author[pac]{N.~Davis}
\ead{Nikolaos.Davis@ifj.edu.pl}

\author[uoa]{F.~K.~Diakonos}
\ead{fdiakono@phys.uoa.gr}

\author[uoa]{G.~Doultsinos}
\ead{georgedoultsinos@gmail.com}

\author[uoa]{N.~Kalntis}
\ead{nikos.kalntis@gmail.com}

\author[uoa]{A.~Kanargias}
\ead{akanaryias@hotmail.com}

\author[uoa]{A.~S.~Kapoyannis}
\ead{akapog@phys.uoa.gr}

\author[pac]{V.~Ozvenchuk}
\ead{Vitalii.Ozvenchuk@ifj.edu.pl}

\author[uoa,sin]{C.~N.~Papanicolas}
\ead{cnp@phys.uoa.gr~;~president@cyi.ac.cy}

\author[pac] {A.~Rybicki}
\ead{Andrzej.Rybicki@ifj.edu.pl}

\author[uoa]{E.~Stiliaris}
\ead{stiliaris@phys.uoa.gr}

\cortext[dec]{deceased}

\address[uoa]{Faculty of Physics, University of Athens, GR-15784 Athens, Greece}
\address[pac]{Institute of Nuclear Physics, Polish Academy of Sciences, Krak\'ow, Poland}
\address[sin]{The Cyprus Institute, Nicosia, Cyprus}

\begin{abstract}
In a systematic search for the QCD critical point in nuclear collisions, at the CERN SPS, it was found that intermittency measurements in the freeze-out state of central \SiSi collisions, at the maximum SPS energy, provide us with an indication of sizeable critical fluctuations. Also, rather recently, a weaker effect was traced in preliminary data of the \ArSc reaction for \crange{10}{20} most central collisions at (approximately) the same energy. However, the uncertainties in the analysis and the limitations of the experimental event statistics make the interpretation of the above measurements (\NAFortyNine, \mbox{\NASixtyOne}) rather inconclusive, inviting for a further, phenomenological investigation with complementary tools and theoretical ideas. To this end, in the present work, we employ intermittency techniques within a model-independent analysis scheme (AMIAS), a novel method from Data Science \cite{AMIAS_cnp}, in order to produce unbiased results for the parameters of the power-laws and in particular for the associated power-law exponent (intermittency index) $\phi_2$. Using data-sets at different peripheralities, we also study the dependence of the $\phi_2$-value on the number of wounded nucleons, in order to uncover the approach to the critical point. With these findings and the help of Ising-QCD partition function, the interpretation of SPS intermittency measurements and their links to the critical region, are discussed.
\end{abstract}

\begin{keyword}
QCD phase diagram \sep critical region \sep proton intermittency \sep AMIAS
\sep \NASixtyOne experiment
\end{keyword}

\end{frontmatter}

%
% ADDING LINENUMBERS FOR REVIEWING:
%\usepackage{lineno}
%\linenumbers
%

\section{Introduction}
\label{sec:sec1}
During the last decade (2010-2020) a systematic search for QCD critical fluctuations was performed at the CERN SPS (\NAFortyNine, \NASixtyOne) in measurements of intermittency in \AnyReaction collisions \cite{NA49_pions,NA49_protons,NA61_qm_2019}. The aim was to capture critical fluctuations when the chemical freeze out of a particular collision comes close to the critical end point in the QCD phase diagram. This picture relies upon the experimental fact that chemical freeze out in \AnyReaction collisions is a state in thermal equilibrium and, therefore, universal fluctuations at the critical point, dependent on the static critical exponents, may, in principle, be revealed. In fact, intermittency in transverse momentum space, linked to the order parameter of the critical point ($\sigma$-field simulated by $\pi^+ \pi^-$ pairs or net-baryon number) is a manifestation of finite size scaling, a fundamental property of critical systems \cite{Ortmanns,Antoniou_prd_2018}. The formulation of intermittency along these lines leads to the following prediction at the critical point, for the two choices of the order parameter \cite{Diakonos_CPOD_2006,Antoniou_prl_1998}:
\begin{enumerate}[(a)]
\item density of $\pi^+ \pi^-$ with zero effective mass ($\sigma$-field) $\Rightarrow$\\ $\Delta F_2 \sim \left( M^2 \right)^{\displaystyle{\phi_2^{(\sigma)}}}~~;$ $~~\displaystyle{\phi_2^{(\sigma)}}=1-\displaystyle{\frac{2 \beta}{3 \nu}}$,
\item baryon-number density $\Rightarrow$ $\Delta F_2 \sim \left( M^2 \right)^{\displaystyle{\phi_2^{(b)}}}~~;~~\displaystyle{\phi_2^{(b)}}=1-\displaystyle{\frac{\beta}{3 \nu}}$
\end{enumerate}
where $\Delta F_2$ is a properly defined correlator, associated with scaled factorial moment $F_2$ \cite{Davis_PhD}, $\beta \approx \displaystyle{{1 \over 3}}$, $\nu \approx \displaystyle{{2 \over 3}}$ are 3d Ising critical exponents and $M^2$ the number of 2d bins in transverse momentum space.

Within this general framework, in a series of measurements performed by the SPS \NAFortyNine experiment, an analysis of nuclear reactions \CC, \SiSi and \PbPb at the maximum SPS energy concluded that:
\begin{enumerate}[(a)]
\item there is no intermittency effect in the processes \CC and \PbPb, whereas
\item a sizeable intermittency effect with critical characteristics is observed in \SiSi, at maximum SPS energy with $\displaystyle{\phi_2^{(\sigma)}}\approx 0.35$, $\displaystyle{\phi_2^{(b)}} = 0.96^{+0.38}_{-0.25}$, compared to the QCD prediction
$\displaystyle{\phi_2^{(\sigma)}}=\displaystyle{{2 \over 3}}$, $\displaystyle{\phi_2^{(b)}}=\displaystyle{{5 \over 6}}$ \cite{NA49_pions,NA49_protons}.
\end{enumerate}

In the successor to \NAFortyNine, the SPS-\NASixtyOne experiment, measurements of proton intermittency in the reactions \BeBe, \ArSc at maximum SPS energy (\cms = 16.8~GeV) lead to the following preliminary results \cite{NA61_qm_2019,Davis_CPOD_2017,Davis_PPC_2019}: there is no intermittency effect either in \BeBe, or in \ArSc for extremely central collisions (\crange{0}{5} most central) of the \ArSc reaction. However, a non zero effect was traced in peripherality \crange{10}{20} of the same reaction, \cite{NA61_qm_2019,Davis_PPC_2019} reflecting possibly a small increase of freeze-out temperature with increasing peripherality due to a decrease of the fireball size \cite{Becattini_2006,Becattini_2014}, bringing, presumably, the system closer to the critical region. Nevertheless, the interpretation of this result remains inconclusive since no reliable estimate of the exponent $\displaystyle{\phi_2^{(b)}}$ can be fixed, mainly due to the uncertainties induced by correlated bins and large statistical errors in the intermittency treatment \cite{Davis_PPC_2019,Michael_corrfit}. As a result, we notice that there exists a valuable compilation of SPS intermittency measurements which show a critical activity in the region of nuclear sizes A=30-40 at the maximum SPS energy of various \AnyReaction collisions, inconclusively interpretable due to the aforementioned uncertainties. In the present work, we make an attempt to understand the significance of these measurements in a phenomenological treatment, with complementary tools and theoretical ideas.

We employ a model-independent analysis scheme (AMIAS) in order to produce unbiased results for the parameters of the power laws and overcome the obstacle of correlated bins in the intermittency treatment \cite{AMIAS_cnp,Michael_corrfit}. We also make use of the Ising-QCD partition function, close to the critical point \cite{Antoniou_prd_2018}, in order to discuss the location of the singularity and interpret the measurements in \ArSc at different centralities. Our description, along these lines, is organized as follows:
In section~\ref{sec:sec2}, the general principles of the AMIAS method are presented, emphasizing the treatment of correlated data. In section~\ref{sec:sec3}, the AMIAS results for the intermittency parameters, limited to the measurements in \SiSi (\crange{0}{12}) and \ArSc for varying centralities, are discussed. Also, the plot of $\phi_2$ versus $N_w$ (number of wounded nucleons) is constructed \cite{Pulawski_2018}, illustrating the approach of \NASixtyOne freeze-out states, towards the critical point. In section~\ref{sec:sec4}, the location of the SPS freeze-out states (\NAFortyNine, \NASixtyOne) in the phase diagram is investigated with respect to the critical region. In particular, it is examined whether the AMIAS intermittent solutions are accommodated within the critical region or remain outside but close to the boundaries.  This study becomes instrumental for the experimental estimate of the size of the critical region. Theoretically, the Ising-QCD partition function predicts a critical region that is a few MeV wide along the chemical potential direction \cite{Antoniou_prd_2018}. Finally in section~\ref{sec:sec5}, our results and conclusions are summarized.

\section{The AMIAS method}
\label{sec:sec2}

The AMIAS method \cite{AMIAS_cnp, AMIAS_stiliaris} has been developed in order to extract physical information from experimental or simulated data with the highest possible
precision and in an unbiased way. It is based on statistical concepts and is able to handle a rather large number of
parameters using Monte Carlo techniques. The method requires the definition of a theory or model which links the parameters
to be determined in an explicit way with the data. Although AMIAS is well suited to resolve physical parameters from data
where the underlying model cannot be inverted, it can equally well address simple fit cases with noisy data.
Recently, AMIAS has been successfully applied in the analysis of pion photoproduction data for the extraction of the multipole excitation amplitudes \cite{AMIAS_Photo}
as well as in lattice QCD investigations for Tetraquark interpolating fields \cite{AMIAS_D0}.

For a given set of parameters that defines the physical model, the principal idea behind AMIAS is that any arbitrary
value assigned to the parameters constitutes a possible solution. Each of these solutions is weighted with a probability value
retrieved out of the data set by a cost function. The information obtained from the given set of data is a Probability Density Function (PDF) assigned
to each of the model parameters. The central values of the parameters and their uncertainties are therefore the expectation values
and the standard deviations of the corresponding PDFs. The sampling method used in AMIAS allows all fit parameters to randomly vary and to yield solutions with all
allowed values, including the insensitive exponential terms.

In the present work our aim is to use the AMIAS protocol for the detection of power-law behaviour in the density-density correlation function of protons in transverse momentum space for small transverse momentum differences. This scenario is expected to be realized in protons produced in the central rapidity region in ion collisions, whenever the produced fireball freezes-out at the thermodynamic conditions (baryochemical potential, temperature) of the QCD critical point \cite{Antoniou_prl_1998}. In fact, this effect is the momentum space counterpart of finite size scaling in configuration space \cite{Antoniou_jopg_2019} revealed through the Fourier transform of the real space density-density correlation function \cite{Antoniou_PRC_2016}. The relevant information is contained in the second order factorial moment $F_2$ counting the, properly normalized, mean number of proton pairs per transverse momentum space cell \cite{Bialasz_1986}. A given transverse momentum domain is partitioned into $M^2$ cells and $F_2$ is calculated for different values of $M$. The presence of power-law density-density correlations is reflected in the function $F_2(M)$ which in this case behaves, in the limit of large $M$, as $\sim M^{2 \phi_2}$ where $\phi_2$ is the intermittency index and the associated physical scaling effect is characterized as intermittency. In experimental data, the intermittency effect may be masked by the presence of noise, i.e. proton pairs which do not follow a power-law, or particles misidentified as protons.
In addition, the fact that the values of $F_2(M)$ for different $M$s are correlated raises the question how these correlations affect the corresponding errors which are unavoidably present due to finite statistics. Before applying AMIAS to experimental data in the next section, we demonstrate here its power to overcome these difficulties.

For this purpose, we use a modified version of the Critical Monte-Carlo (CMC) event generator \cite{CMC,CMCb} suited for simulating protons in transverse momentum space; simulated protons are produced through a truncated L\'evy walk process to exhibit density-density correlations mimicking those originating from a fireball freezing out at the QCD critical point. The power-law exponent is chosen to describe correlations characterizing a critical system in the 3d-Ising universality class; associated intermittency index has the value $\phi_2={5 \over 6}$ \cite{Antoniou_prl_1998}. Furthermore, the algorithm can be parametrized \cite{Davis_CPOD_2017} to produce an exponential one-particle proton $p_T$ distribution and a Poissonian per-event proton multiplicity distribution, the values of which can be plugged in; finally, truncated L\'evy walk bounds can be fine-tuned in order to produce critical density-density correlations within the desired scales.

A number of uncorrelated proton momenta drawn from a one-particle $p_T$ distribution are interspersed among the critical protons at an adjustable percentage. These simulate the effect of non-critical background contamination on the critical signal.

In the following text, and unless stated otherwise, proton intermittency analysis of $F_2(M)$ is applied within a rectangular region in transverse momentum space: $-1.5 \leq p_x, p_y \leq 1.5$~GeV/$c$. The number $M$ of one-dimensional bins in momentum space ranges from 1 to 150. The power-law fit for the intermittency index $\phi_2$ is performed in the $M^2 \in [10^3 , 10^4]$, or equivalently the $M \in [32 , 100]$, range.

\begin{figure}[htbp!]
\begin{center}
\raisebox{-0.05\height}{\includegraphics[width=0.32\textheight]{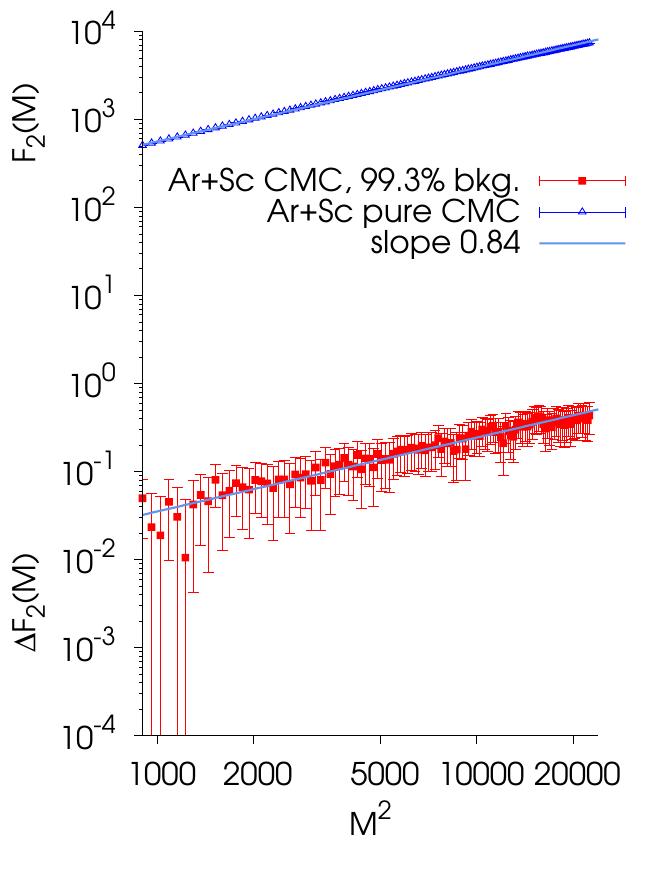}}
\raisebox{0.0\height}{\includegraphics[width=0.38\textheight]{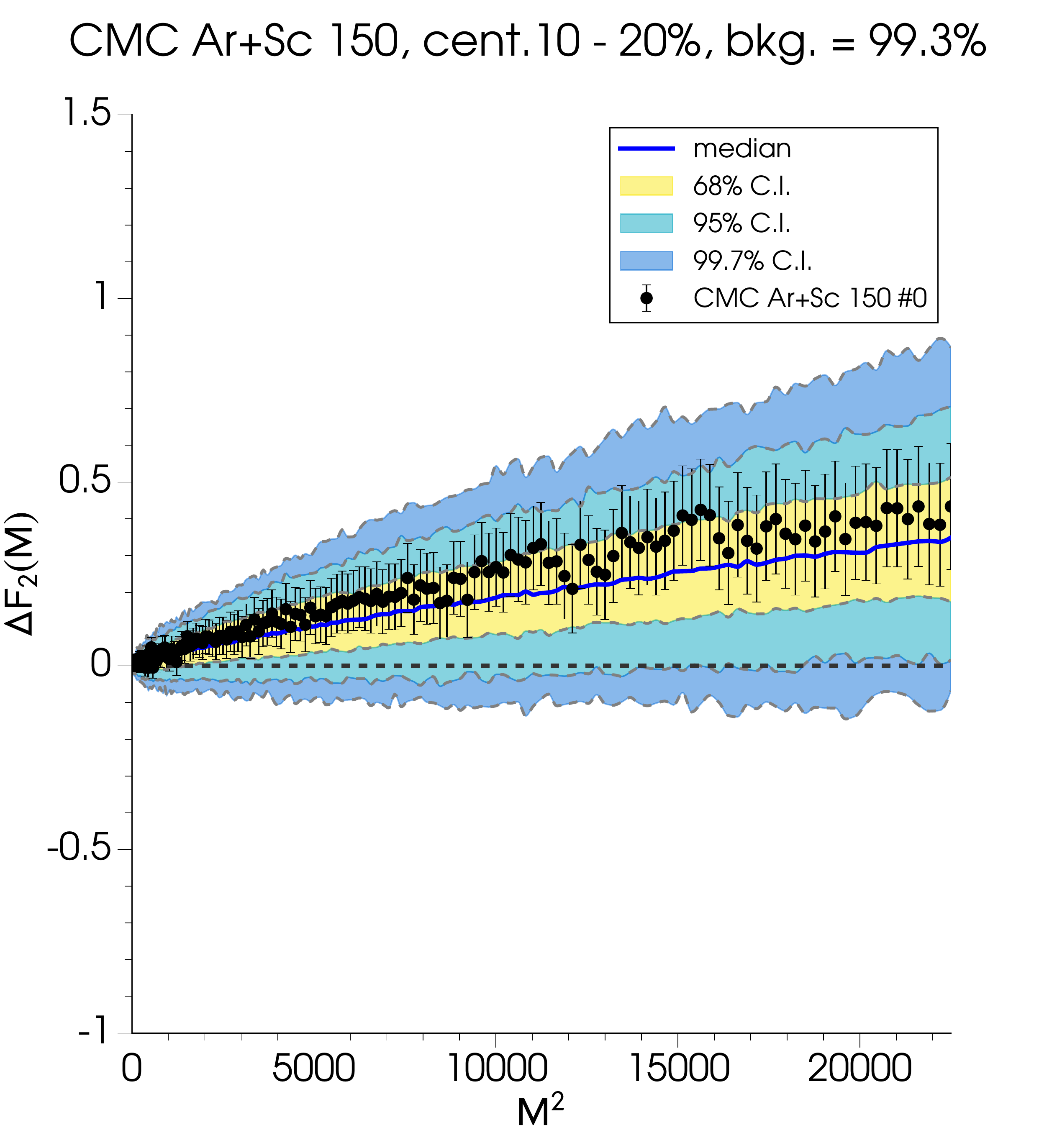}}
\end{center}
\caption{(Left) $F_2(M)$ of pure (blue triangles) and $\Delta F_2(M)$ of background-diluted CMC (red squares) \ArSc simulated collisions (\crange{10}{20} most central,$\sqrt{s_{NN}}=16.8$~GeV). Slope lines of $\phi_{2,cr} \simeq 0.84$ are fitted to both sets to guide the eye; (Right) $\Delta F_2(M)$ of background-diluted CMC for the same $\sim 400$K event CMC iteration (black points \& error bars), along with $\Delta F_2(M)$ confidence intervals (68-95-99.7\%) for 400 independent iterations.}
\label{fig:ArScCMC}
\end{figure}

For our AMIAS test, we generate sets of $\sim 400$K proton events with the characteristics of the \NASixtyOne \ArSc \crange{10}{20} most central data set at 150$A$~GeV/$c$ beam momentum ($\cms =16.8$~GeV). Average proton multiplicity per event is adjusted to $\sim 2.5$, with a standard deviation of $\sim 1.4$. Critical proton tracks are contaminated with $99.3 \%$ noise ($0.7\%$ critical component). This huge noise masks the underlying power-law behaviour at the level of $F_2(M)$. As shown in \cite{NA49_protons} in this regime, the power-law component is revealed in the correlator $\Delta F_2(M) \equiv F_{2,\mathrm{data}}(M)-F_{2,\mathrm{mixed}}(M)$, accounting for the subtraction of the background simulated by mixed events, $F_{2,\mathrm{mixed}}(M)$. This behavior is evident in Fig.\ref{fig:ArScCMC} (left), where the pure and background-diluted CMC factorial moments are shown in double logarithmic scale: the correlator $\Delta F_2(M)$ retains the critical slope of the pure CMC $F_2(M)$, even though its values are reduced by orders of magnitude compared to pure CMC. Fig.\ref{fig:ArScCMC} (right) shows $\Delta F_2(M)$ for the same set of events, plotted against the distribution of $\Delta F_2(M)$ values for a total of 400 independent CMC sets, under the same conditions (linear scale is used to accommodate negative values). One observes that the original sample is typical, falling mostly within the central yellow band roughly corresponding to 1$\sigma$, whereas a minority of samples have either lower or higher slopes. However, in itself the distribution of $\Delta F_2(M)$ values does not give a complete picture of the power-law behaviour of moments due to the role of bin correlations.

The above generated sets have been analyzed with the AMIAS method. A total of 400 independent iterations with 150 data points each are assembled in a grand examination set.
The function used to fit the data had the following analytical form
\begin{equation}
\Delta F_2(M) = 10^{a_0} \left( \frac{M^2}{10^4} \right) ^{\phi_2}
\label{eq:power_law_AMIAS}
\end{equation}
where $a_0$ and $\phi_2$ are the two fit parameters. The uniform sampling method has been applied in the Monte-Carlo technique choosing appropriate variation range for both parameters.
Millions of possible solutions are selected by eliminating extreme values with the help of a $\chi^2$-cut. The final PDF solutions for the fit parameters are constructed in the
usual way by weighting each point of the accumulated sample with the factor $e^{-\chi^2/2}$.

\begin{figure}[htbp!]
\begin{center}
\raisebox{-0.05\height}{\includegraphics[width=0.70\textheight]{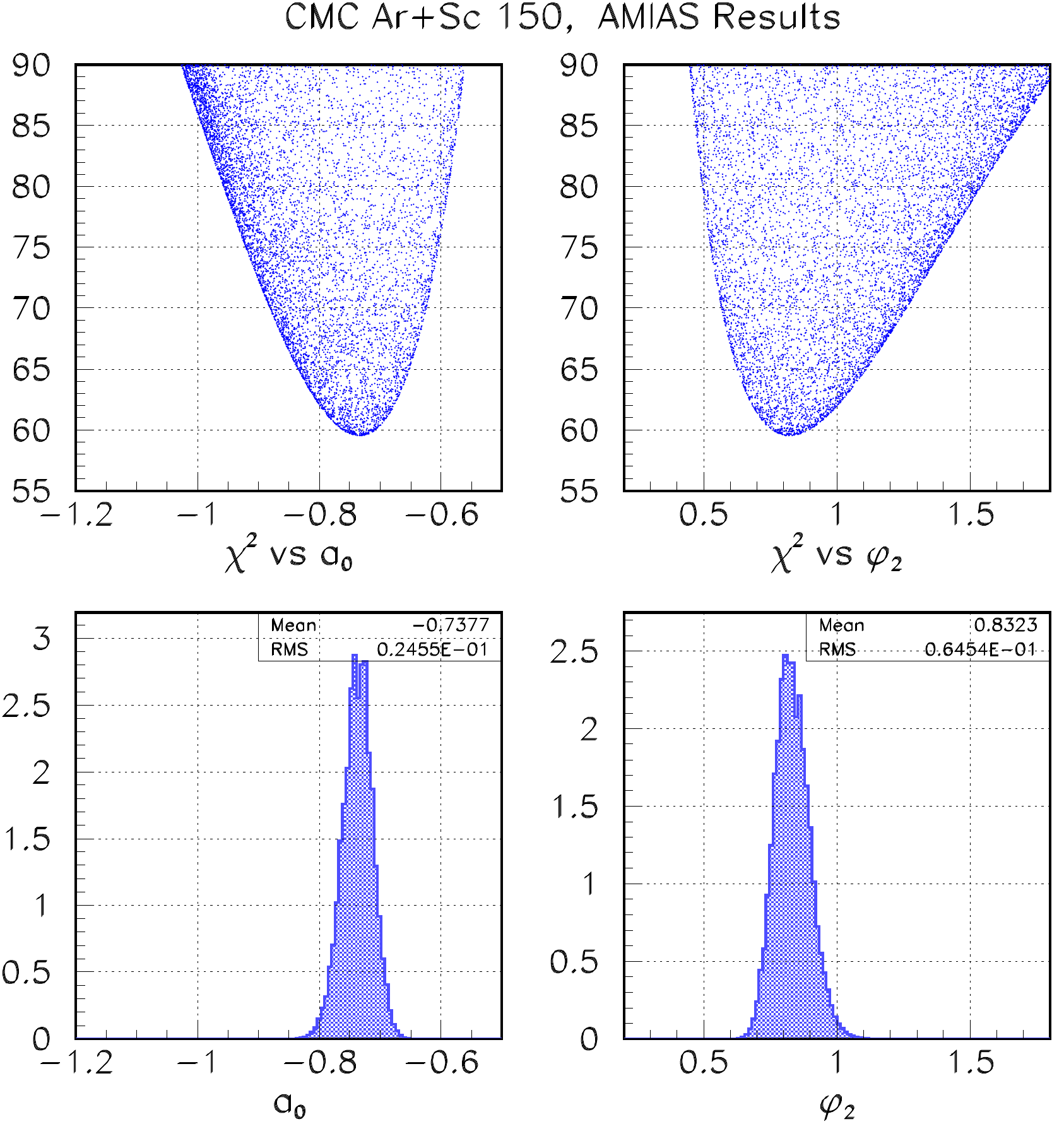}}
\end{center}
\caption{AMIAS results for the two fit parameters $a_0$ and $\phi_2$ from the CMC generated 400 independent iterations. Top row (scatter plot) shows
the microcanonical solution for each parameter. The histograms in the bottom row represent the non-normalized PDF solutions for both fit parameters.}
\label{fig:AMIAS_CMC}
\end{figure}

\begin{figure}[htbp!]
\begin{center}
\raisebox{-0.05\height}{\includegraphics[width=0.50\textheight]{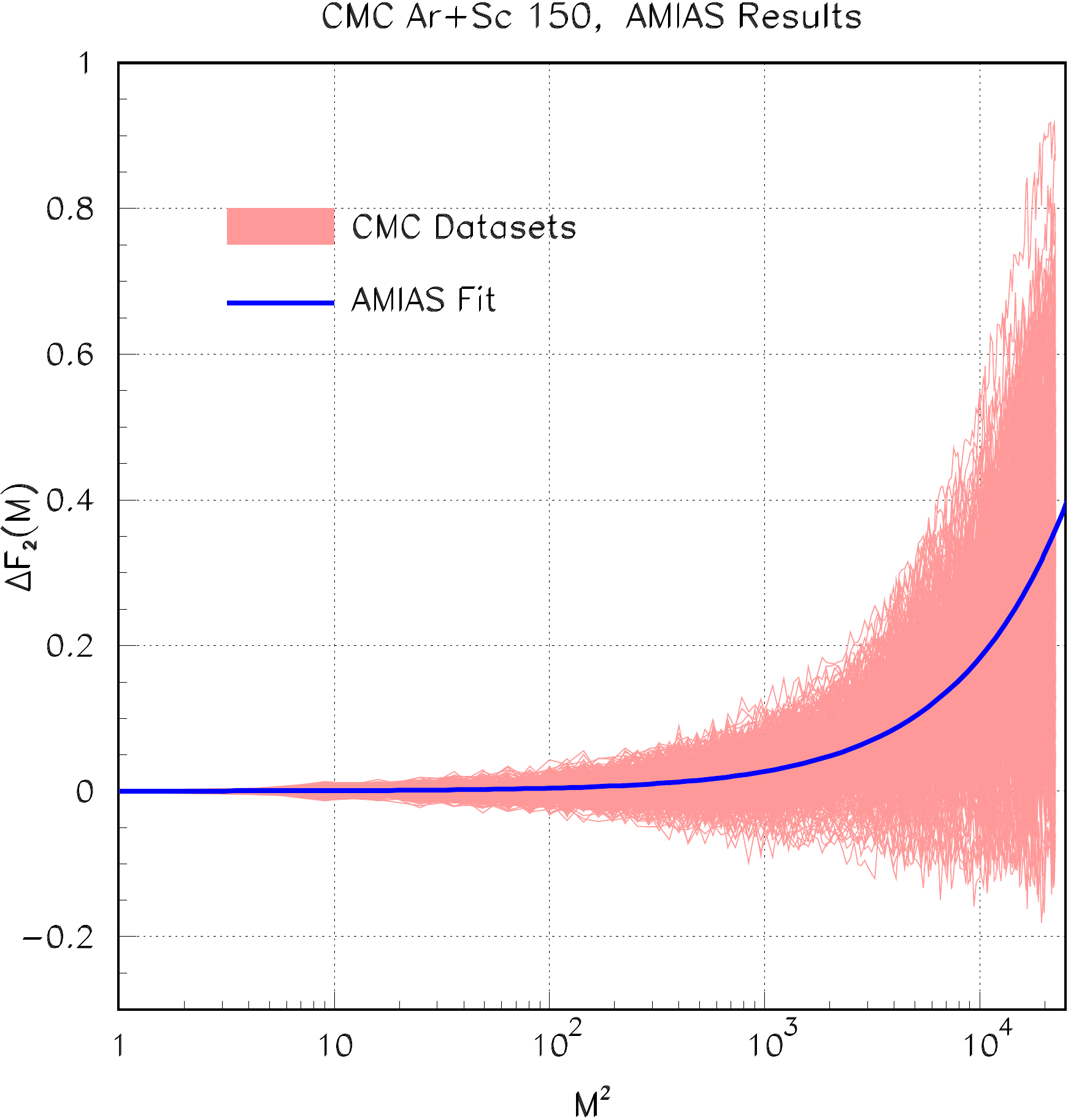}}
\end{center}
\caption{The AMIAS solution (line) extracted from the central parameter values together with all generated CMC data sets represented here as a colored band.}
\label{fig:AMIAS_CMC_fit}
\end{figure}

The obtained PDFs for the two parameters $a_0$ and $\phi_2$ are shown in Fig.~\ref{fig:AMIAS_CMC} together with the microcanonical ensemble of both solutions.
The extracted value for the power law parameter is $\phi_2 = 0.83 \pm 0.06$, which is in perfect agreement with the value used in the generator.
Fig.~\ref{fig:AMIAS_CMC_fit} summarizes all generated CMC data sets with the AMIAS solution extracted from the central parameter values.

In the following, we will focus on the extraction of the power-law behaviour in $\Delta F_2(M)$ obtained from experimental data in \NAFortyNine and \NASixtyOne experiments.

\section{Intermittent fluctuations in SPS (CERN)}
\label{sec:sec3}

%Intermittency analysis provides a valuable tool in the search for critical fluctuations based on first principles. As mentioned in the introduction, finite-size scaling, which is an indicator of the proximity to the critical point, leads to singular behaviour of the momentum space density-density correlation function for small momentum differences. This singularity reflects the formation of clusters with fractal geometry detectable through the power-law behaviour of the associated second factorial moment.

\begin{figure}[htbp]
\includegraphics[width=\textwidth]{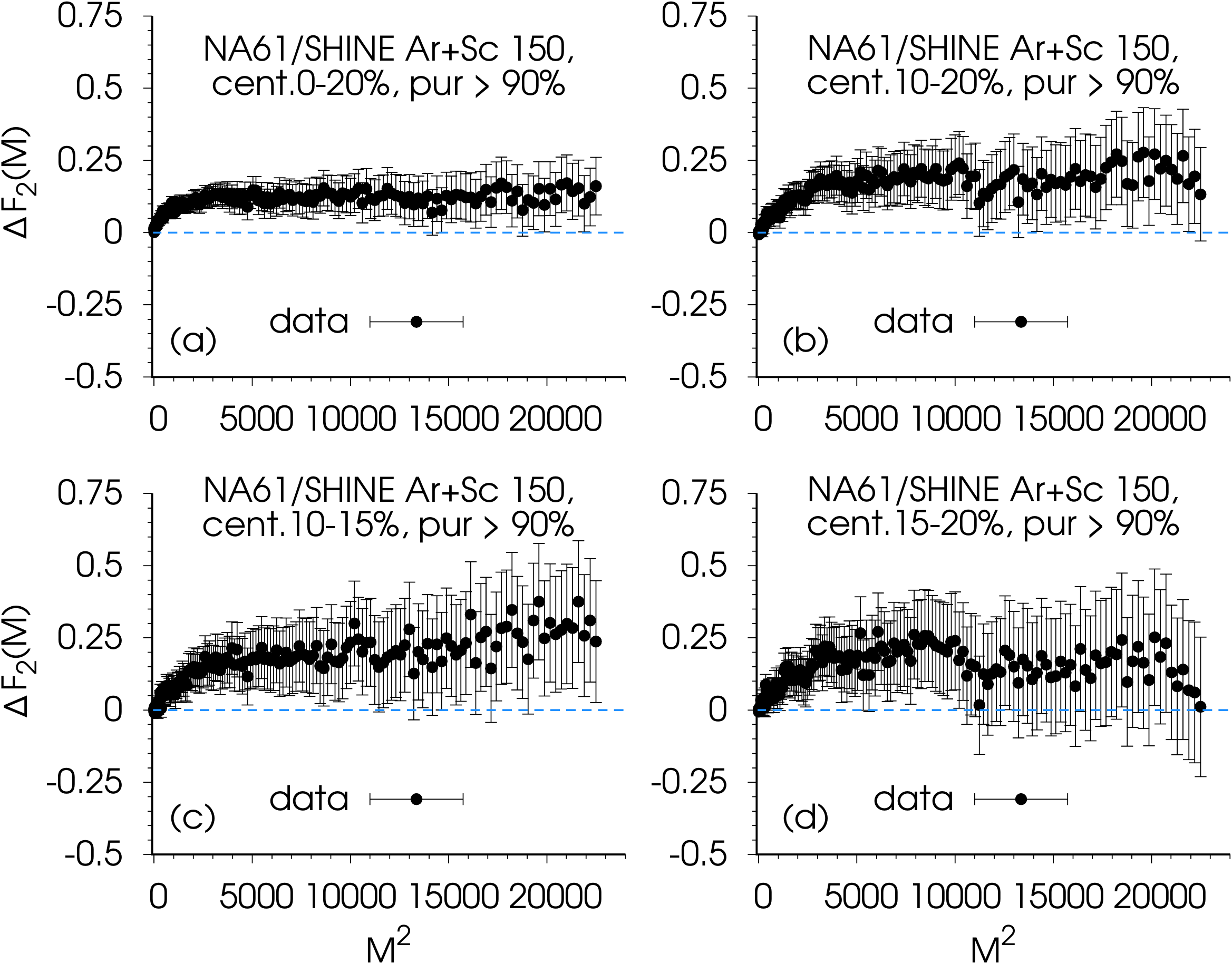}
\caption{$\Delta F_2(M)$ for \NASixtyOne \ArSc\ {\em (a)} \crange{0}{20}, {\em (b)} \crange{10}{20}, {\em (c)} \crange{10}{15}, and {\em (d)} \crange{15}{20} most central collisions at 150$A$~GeV/$c$. Error bars correspond to statistical errors estimated via the bootstrap method \cite{Metzger,Efron-Hesterberg}.}
\label{fig:DF2_ArSc150_abcd}
\end{figure}

In this section we employ AMIAS to estimate the intermittency index $\phi_2$ based on the \NASixtyOne (CERN, SPS) preliminary measurements of the proton transverse momentum space correlator $\Delta F_2$ in \ArSc collisions at 150A~GeV/$c$ beam momentum for different peripheralities in the range \crange{0}{20} \cite{Davis_PPC_2019}. Our goal is to explore how the associated $\phi_2$ distribution, determined by AMIAS, depends on peripherality of the collisions. As mentioned also in the Introduction, such a study is twofold motivated:
\begin{itemize}
\item Freeze-out states produced in \ArSc collisions at 150A~GeV/$c$ beam momentum are expected to lie, in the baryochemical potential ($\mu_B$) and temperature ($T$) plane, relatively close to $(\mu_{B,Si},T_{Si})$, parametrizing the freeze-out state produced in central \SiSi collisions at 158A~GeV/$c$ (\NAFortyNine experiment at CERN-SPS). In the latter system intermittent fluctuations with critical characteristics have been observed \cite{NA49_pions,NA49_protons}.
\item The theoretical prediction that the critical region is very narrow in $(\mu_B,T)$ plane \cite{Antoniou_jopg_2019}, implies that the neighbourhood of $(\mu_{B,Si},T_{Si})$ should be explored with small steps in $\mu_B$ and $T$.  Such a fine search can be realized in a suitable scan, employing freeze-out states of \ArSc collisions at 150A~GeV/$c$ at different peripheralities. This is due to the fact that there is experimental evidence as well as theoretical understanding \cite{Becattini_2006,Becattini_2014} that changes in the peripherality influence the freeze-out conditions $(\mu_B,T)$ in a prescribed mild manner.
\end{itemize}
In fact it is expected that more peripheral \ArSc collisions can approach the freeze-out conditions of central \SiSi collisions at similar energy. Unfortunately, experimental limitations restrict the range of peripherality variation in the interval \crange{0}{20}. Within these constraints the \NASixtyOne experiment has measured the correlator $\Delta F_2$ for \ArSc collisions at 150A~GeV/$c$ in different peripherality zones of increasing resolution. Starting from the coarsest scale \crange{0}{20}, the \crange{0}{20} zone is subsequently partitioned into two non-overlapping intervals of $10\%$ width, i.e. \crange{0}{10} and \crange{10}{20}, and the resolution is further increased though a subdivision into four non-overlapping intervals of $5\%$ width each: \crange{0}{5}, \crange{5}{10}, \crange{10}{15} and \crange{15}{20} providing the finest scale in this analysis. In Fig.~\ref{fig:DF2_ArSc150_abcd}(a-d)  we present, indicatively, $\Delta F_2(M)$ plots for the peripheralities: (a) \crange{0}{20}, (b) \crange{10}{20}, (c) \crange{10}{15} and (d) \crange{15}{20}. The remaining $\Delta F_2$ measurements at peripheralities \crange{0}{10}, \crange{0}{5} and \crange{5}{10} are not shown here; we refer the interested reader to \cite{Davis_PPC_2019} for a detailed presentation.

\begin{figure}[h]
\begin{center}
\raisebox{-0.05\height}{\includegraphics[width=0.55\textheight]{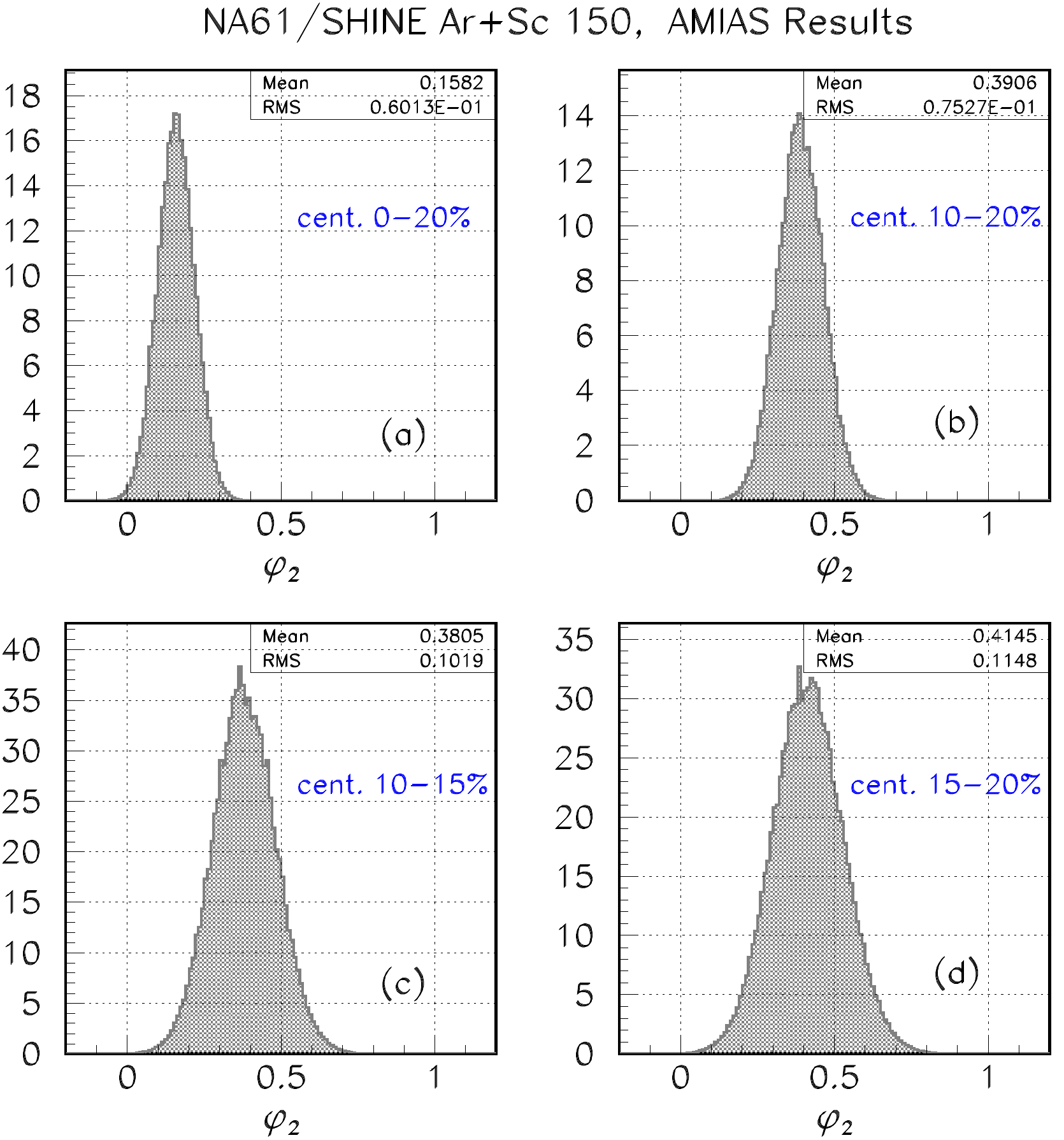}}
\end{center}
\caption{AMIAS analysis results for the $\phi_2$-distribution for the corresponding cases for $\Delta F_2$ shown in Fig.~\ref{fig:DF2_ArSc150_abcd}(a-d).}
\label{fig:DF2_AMIAS_abcd}
\end{figure}

\begin{figure}[h]
\begin{center}
\raisebox{-0.05\height}{\includegraphics[width=0.55\textheight]{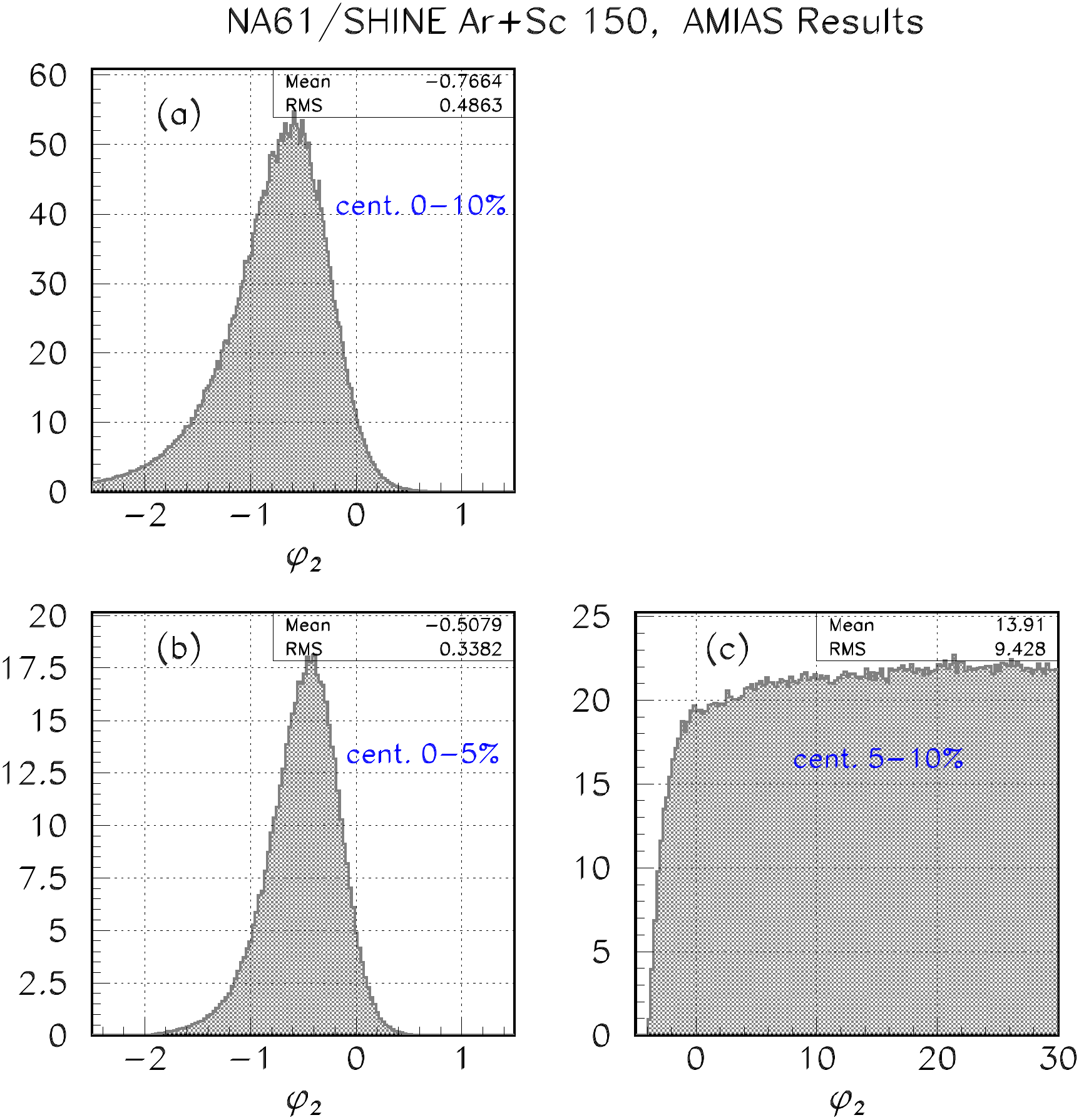}}
\end{center}
\caption{AMIAS analysis results for the $\phi_2$-distribution corresponding to the second set of \NASixtyOne $\Delta F_2$ results
concerning the \crange{0}{10}, \crange{0}{5} and \crange{5}{10} peripherality zones.}
\label{fig:DF2_AMIAS_abc}
\end{figure}

We use the AMIAS protocol to estimate the distribution of the intermittency index $\phi_2$ for each of the cases described above. The results of this analysis are presented in two figures. In Fig.~\ref{fig:DF2_AMIAS_abcd}(a-d) we present the $\phi_2$-distribution for the corresponding cases for $\Delta F_2$ shown in Fig.~\ref{fig:DF2_ArSc150_abcd}(a-d) while in Fig.~\ref{fig:DF2_AMIAS_abc}(a-c) we show the $\phi_2$-distribution corresponding to the second set of \NASixtyOne $\Delta F_2$ results concerning the \crange{0}{10}, \crange{0}{5} and \crange{5}{10} peripherality zones respectively.

We observe in Figs.~\ref{fig:DF2_AMIAS_abc}(a-c) that no intermittency effect is detected in the peripherality bins \crange{0}{10} (and consequently also in the subdivisions \crange{0}{5} and \crange{5}{10}) while a clear intemittency effect, expressed by a non-vanishing mean $\phi_2$ value and a standard deviation sufficiently small to exclude the $\phi_2 \leq 0$ region, is observed in the peripherality bin \crange{10}{20}, as well as its subdivisions \crange{10}{15} and \crange{15}{20}, shown in Figs.~\ref{fig:DF2_AMIAS_abcd}(a-d). The mean value $\langle \phi_2 \rangle$ is slightly increased in the \crange{15}{20} as compared to that in the \crange{10}{15} peripherality interval. On average, in the \crange{0}{20} zone, a small but certainly non-vanishing intermittency effect is sustained. These results are compatible with a tendency of the \ArSc freeze-out state to approach the \SiSi freeze-out state with increasing peripherality. This behaviour was theoretically foreseen in \cite{Antoniou_jopg_2019}.

\begin{figure}[htbp]
\includegraphics[width=0.475\textwidth]{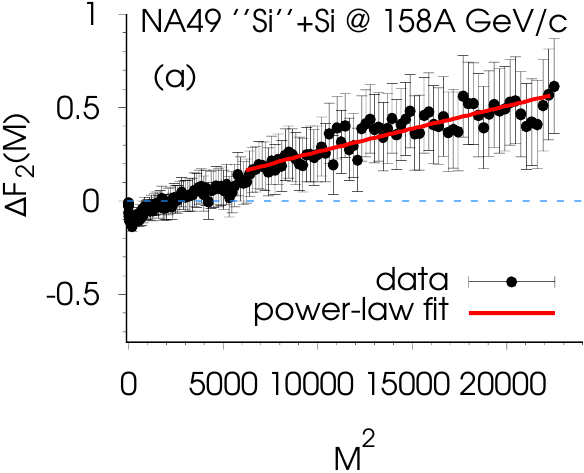}
\includegraphics[width=0.400\textwidth]{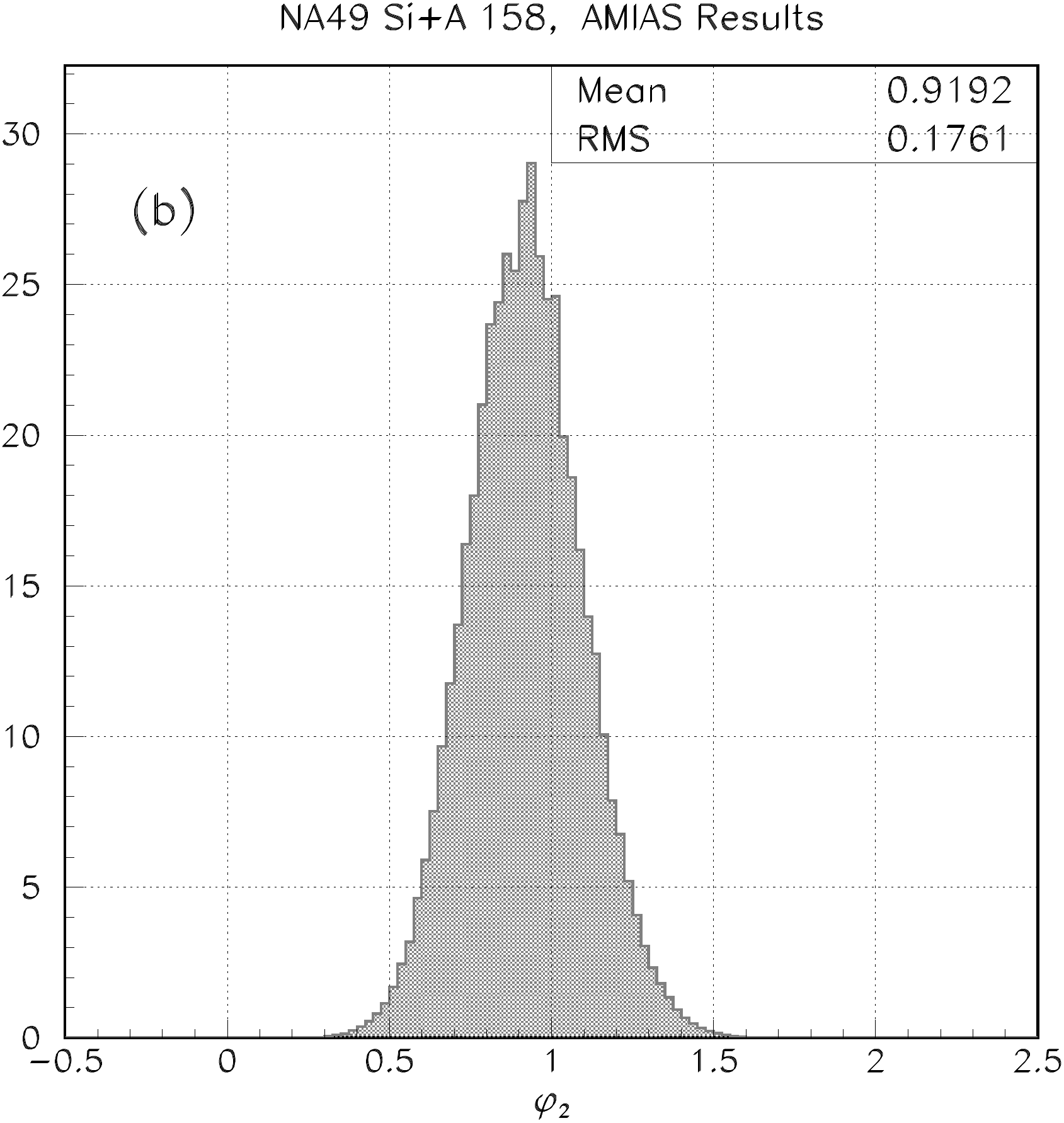}
\caption{{\em Left:} $\Delta F_2(M)$ for \NAFortyNine \SiSi 0-12\% most central collisions at 158$A$~GeV/$c$ \cite{NA49_protons}.
{\em Right:} AMIAS distribution for the intermittency index $\phi_2$.}
\label{fig:DF2_NA49_Si}
\end{figure}

For completeness we have used AMIAS to determine the distribution of the intermittency index $\phi_2$ also for the \SiSi system analysed in detail in \cite{NA49_protons}. In Fig.~\ref{fig:DF2_NA49_Si}a we show the correlator - for protons in transverse momentum space - $\Delta F_2(M)$ (\SiSi collisions, peripherality \crange{0}{12}, 158A~GeV/$c$) as measured in the \NAFortyNine experiment (SPS, CERN), while in Fig.~\ref{fig:DF2_NA49_Si}b, we display the $\phi_2$ distribution obtained with AMIAS. For consistency, we have used in AMIAS the same fitted $M^2$-range as in \cite{NA49_protons}, $M^2 \in [6000, 22500]$. The overall form of the distribution is compatible with the result published in \cite{NA49_protons}. The advantage of the AMIAS method is that it leads to a significantly narrower distribution than that obtained by applying the bootstrap method \cite{Metzger,Efron-Hesterberg} to  $\phi_2$ fits, as detailed in \cite{NA49_protons}.

A transparent interpretation of the results shown in Figs.~(\ref{fig:DF2_AMIAS_abcd}-\ref{fig:DF2_NA49_Si}) is possible if one uses the number of wounded nucleons $N_w$ for the characterization of the freeze-out states associated with each peripherality bin.

The mean number of wounded nucleons in each bin of \ArSc collisions is estimated by means of a geometrical Glauber simulation~\cite{glauber}. The Ar nuclear density profile is obtained from the Fourier-Bessel expansion of the nuclear charge density~\cite{vries} while the Sc profile is taken as a Saxon-Woods distribution with half density and surface thickness terms taken as 0.542~fm and 3.77~fm, respectively~\cite{vries,angeli}. The nucleon-nucleon cross-section at this energy is taken as 31.42~mb in agreement with literature~\cite{pp}. In order to match the specific \NASixtyOne centrality selection based on the energy deposit in a forward calorimeter~\cite{jinst}, the simulated centrality samples are defined by the energy of projectile (Ar) spectator nucleons. The potential influence of the additional energy deposit by produced charged particles in the \NASixtyOne forward calorimeter can be deduced from Ref.~\cite{kiel} and it is included in the uncertainty estimated for our calculation, presented in Table~\ref{tab:AMIAS_Nw_phi2}.

In Table~\ref{tab:AMIAS_Nw_phi2} we summarize the results of the AMIAS analysis in the considered systems, presenting the mean value $\langle \phi_2 \rangle$ as well as the corresponding error $\delta \phi_2$ obtained from the $\phi_2$ distributions in Figs.~(\ref{fig:DF2_AMIAS_abcd}-\ref{fig:DF2_NA49_Si}), providing also the estimated $N_w$ value for each analysed data set.
\begin{table}[h]
\begin{center}
\begin{tabular}{| l | c | c | c |}
\hline
Reaction & centrality ($\%$) & $N_w$ & $\langle \phi_2 \rangle$ $(\delta \phi_2)$ \\
\hline
\hline
\ArSc & 0-10 & 62(0.6) & -0.77(49)  \\
\hline
\ArSc & 10-20 & 45.9(0.5) & 0.39(08)   \\
\hline
\ArSc & 0-5 & 66.6(0.9) & -0.51(34)  \\
\hline
\ArSc & 5-10 & 57.3(0.4) & --   \\
\hline
\ArSc & 10-15 & 49.4(0.4) & 0.38(10) \\
\hline
\ArSc & 15-20 & 42.4(0.5) & 0.41(12) \\
\hline
\ArSc & 0-20 & 54(0.6) & 0.16(06) \\
\hline
Si+A & 0-12 & 37(3) & 0.92(18) \\
\hline
\end{tabular}
\end{center}
\caption{AMIAS $\langle\phi_2\rangle$ and corresponding error $\delta\phi_2$ results vs the estimated mean number of wounded nucleons $N_w$ for central \SiSi and different \ArSc peripherality ranges.}
\label{tab:AMIAS_Nw_phi2}
\end{table}

\begin{figure}[h]
\begin{center}
\includegraphics[width=0.8\textwidth]{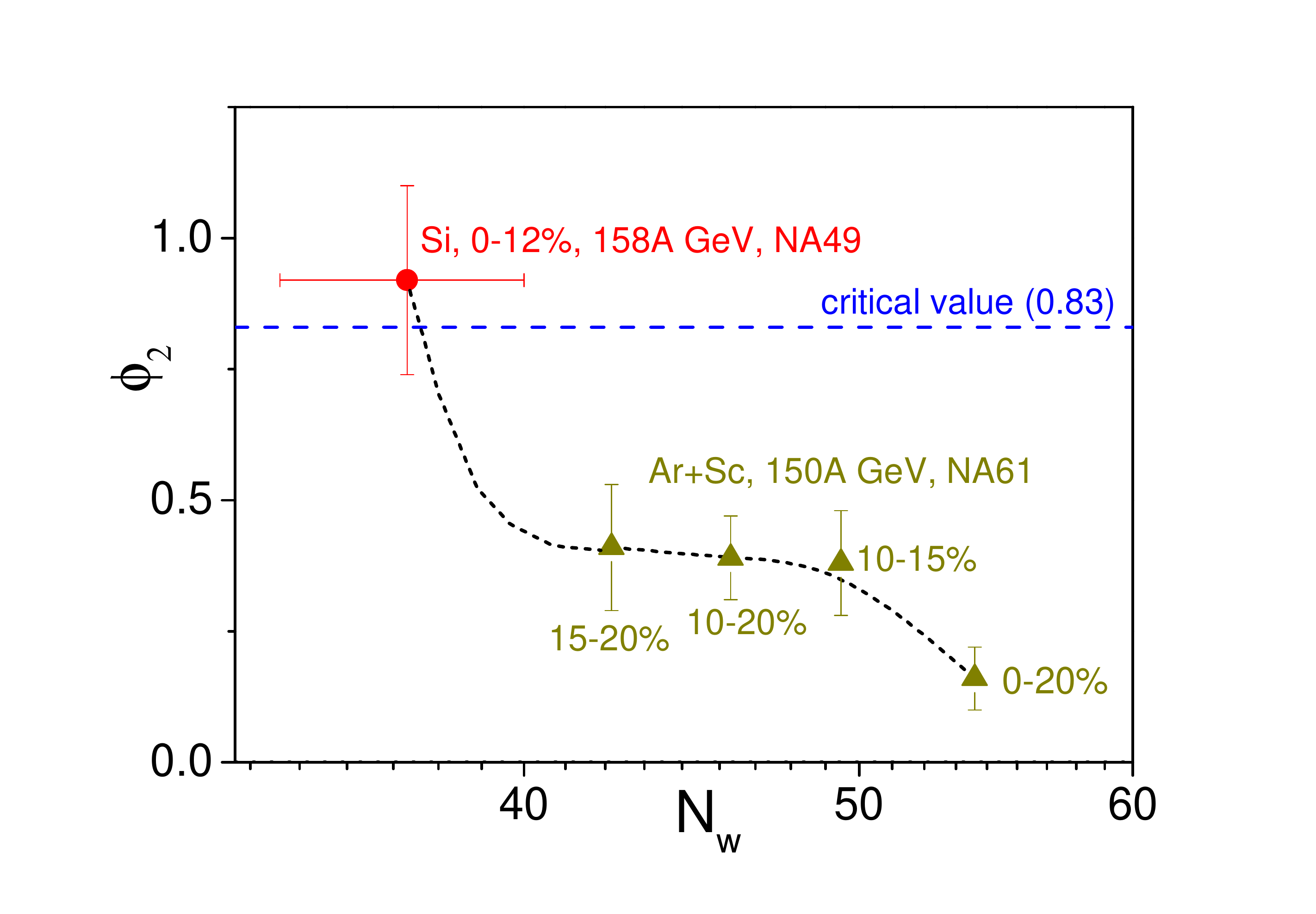}
\caption{The intermittency index $\phi_2$ versus the number of wounded nucleons $N_w$ for the systems: \SiSi at $158$A~GeV/$c$ and peripherality \crange{0}{12} (\NAFortyNine experiment, red circle), \ArSc at $150$A~GeV/$c$ and peripheralities \crange{0}{20}, \crange{10}{20}, \crange{10}{15} and \crange{15}{20} (\NASixtyOne experiment, light green triangles). The dashed blue line indicates the critical $\phi_2$ value $\phi_{2,cr}={5 \over 6}$.}
\label{fig:phi2_vs_Nwounded}
\end{center}
\end{figure}

It is interesting to plot the intermittency index $\phi_2$, calculated with AMIAS, as a function of the number of wounded nucleons, similarly to the plot presented in \cite{Pulawski_2018} using $\phi_2$-values obtained through ordinary fitting procedure. In Fig.~\ref{fig:phi2_vs_Nwounded} we show such a plot including only systems with $N_w < 55$. Since $N_w$ is a quantity related to the effective size of the produced fireball, it is expected to depend smoothly on the freeze-out temperature $T$ \cite{NA49_wounded}. We clearly observe the tendency of the most peripheral \ArSc freeze-out states to approach the \SiSi freeze-out state since $N_w$ is decreasing with increasing peripherality. It is remarkable that, at the same time, the intermittency index $\phi_2$ increases too, indicating the proximity to the critical region. Clearly the $\phi_2$ value for the freeze-out state associated with the most peripheral \ArSc collisions is still far from the critical value ($0.41$ compared to $0.83$) indicating that this system has not entered yet into the critical region \cite{Antoniou_jopg_2019}. Furthermore, the absence of an intermittency effect in the \CC (central collisions at 158A~GeV/$c$, \NAFortyNine) \cite{NA49_protons} and \BeBe (central collisions at 150A~GeV/$c$, \NASixtyOne preliminary) \cite{Davis_CPOD_2017} systems with $N_{w,C} \approx 14 < N_{w,Si} \approx 37$ \cite{NA49_wounded}, implies the formation of a maximum located close to $N_w=N_{w,Si}$ in the $\phi_2(N_w)$ function. Unfortunately, due to very low proton multiplicity and insufficient statistics the $\phi_2$ value for \BeBe is indeterminable. Nevertheless, the overall picture provides us with a rough estimation of the size of the critical region quantified by the differences $\vert \phi_2 - {5 \over 6} \vert$ and $\vert N_w - N_{w,Si} \vert$. A more detailed representation of the critical region and its surroundings requires additional theoretical tools and it will be presented in the next section. There we will show that, in the immediate neighbourhood of the critical point, the quantities $\phi_2$ and $N_w$ form an almost cartesian coordinate system, allowing an alternative parametrization of the QCD phase diagram in this region.

\section{Probing the QCD critical region}
\label{sec:sec4}

In this section we exploit the AMIAS results of the previous section to construct a phase diagram of the critical region and its surroundings along the lines introduced in \cite{Antoniou_jopg_2019}. To this end we employ the Ising-QCD partition function $\mathcal{Z}$, derived in \cite{Antoniou_prd_2018}, to describe the net-baryon density and its fluctuations close to the critical point. To be self-contained, we repeat here the definition of $\mathcal{Z}$:
\begin{equation}
\mathcal{Z}=\displaystyle{\sum_{N=0}^L} \zeta^N \exp\left[-\frac{1}{2} \hat{m}^2 \frac{N^2}{L}-g_4 \hat{m} \frac{N^4}{L^3} - g_6 \frac{N^6}{L^5}\right]
\label{eq:1}
\end{equation}
where $N=N_{B}-N_{\bar{B}}$ is the net-baryon number in volume $V=L \beta_c^3$, $\beta_c=1/k_B T_{cr}$ being the characteristic length scale of the system fixed by the critical temperature $T_{cr}$. The dimensionless variable $L$ quantifies the size of the considered baryonic system. The universal couplings $g_4=0.97 \pm 0.02$, $g_6=2.05 \pm 0.15$ are calculated in \cite{Tsypin1994} while $\hat{m}=\beta_c m$ is related to the correlation length $\xi=m^{-1}$ of the infinite system, given by $\xi=\xi_{0,\pm} \vert t \vert^{-\nu}$ with $\nu \approx {2 \over 3}$ for the 3d Ising universality class and $t=\frac{T-T_{cr}}{T_{cr}}$ the reduced temperature. The index $\pm$ in $\xi_{0,\pm}$ is used to distinguish between the region $T > T_{cr}$ in which the correlation length amplitude is $\xi_{0,+}$ and the region $T < T_{cr}$ where the corresponding amplitude is $\xi_{0,-}$. Notice that the ratio $\frac{\xi_{0,+}}{\xi_{0,-}}$ is fixed and approximately equal to $2$ in the 3d-Ising universality class. Finally, $\zeta=e^{(\mu_B - \mu_{B,cr})\beta_c}$ represents the fugacity in the grand canonical ensemble with $\mu_{B,cr}$ the critical baryochemical potential. As explained in \cite{Antoniou_jopg_2019} Eq.~(\ref{eq:1}) provides a valid thermodynamic description of the critical fluid in the neighbourhood of the critical point where hadronic and quark phases are indiscernible. Furthermore, the partition function (\ref{eq:1}) reproduces accurately all the critical properties, i.e. scaling, critical exponents, of the 3d Ising system \cite{Antoniou_2020}. It can be used to derive scaling laws of the form:
\begin{equation}
\langle N^k \rangle \sim L^{k q}, k=1,~2,...
\label{eq:2}
\end{equation}
expressing the finite size scaling properties of the critical fluid \cite{Antoniou_jopg_2019}. For the 3d-Ising model the power-law exponent is $q \approx {5 \over 6}$. In fact, the partition function (\ref{eq:1}) enables also the description of the baryonic fluid for thermodynamic conditions close, but at a distance from the critical point $(\mu_{B,cr},T_{cr})$. In such a case, scaling laws of the form in Eq.~(\ref{eq:2}) are still approximately valid, however the exponent $q$ is modified to $\tilde{q}$ varying in the range $[0,{5 \over 6})$ or $({5 \over 6},1]$. Then, the difference $\vert \tilde{q} - {5 \over 6} \vert$ provides a measure for the distance from the critical baryochemical potential value $\mu_c$, as shown in \cite{Antoniou_jopg_2019}. Of course, far from the critical point the validity of the partition function $\mathcal{Z}$ in (\ref{eq:1}) breaks down. The importance of the power-law exponent $\tilde{q}$ is that it can be measured in ion collision experiments through intermittency analysis in transverse momentum space. Within this framework, power-law behaviour of the correlator $\Delta F_2(M) \sim M^{2 \phi_2}$ as a function of the number of cells $M$ in transverse momentum space partitioning (intermittency effect), is a manifestation of the finite-size scaling (\ref{eq:2}) in configuration space, both effects related to each other through a Fourier transform \cite{Antoniou_PRC_2016}. As a consequence the intermittency index $\phi_2$ coincides with the finite-size scaling exponent $q$ (or more general $\tilde{q}$).

For the construction of the phase diagram we will use two basic ingredients: (i) the $\phi_2$ values calculated with AMIAS and (ii) the number of wounded nucleons $N_w$ calculated with a geometrical Glauber model simulation \cite{glauber}. For both quantities we will use their central values, given in Table~\ref{tab:AMIAS_Nw_phi2}, neglecting the associated errors. However, the essentials of our treatment, contained in the methodological approach, are in general valid independently of this choice. Initially the phase diagram will be presented in the reduced variables $\ln \zeta$ (reduced baryochemical potential) and $t$ (reduced temperature), however, as we will see in the following, close to the critical point the quantities $\phi_2$ and $N_w$ play a similar role.

Our strategy is as follows:
\begin{itemize}
\item Firstly we determine the curves $\phi_2(\ln \zeta,t)=C_i$ with $i=1,~2...,~5$ and $C_i$ constant values taken from the last column in Table~ \ref{tab:AMIAS_Nw_phi2} in increasing order. We consider only the cases with $C_i > 0$.

\item Subsequently we construct the corresponding $N_w(\ln \zeta,t)=c_i$ curves (with $i=1,..,5$) where the $c_i$ values are taken from the third column in Table~\ref{tab:AMIAS_Nw_phi2}. 

\item The intersections of these families of lines determine the location of the corresponding freeze-out states in the $(\ln \zeta,t)$ plane.
\end{itemize}
%, i.e. $C_1=0.16$ corresponding to \ArSc in \crange{0}{20} peripherality, $C_2=0.38$ (\ArSc, \crange{10}{15} periph.), $C_3=0.39$ (\ArSc, \crange{10}{20} periph.), $C_4=0.41$ (\ArSc, \crange{15}{20} periph.) and $C_5=0.92$ (\SiSi, \crange{0}{12} periph.). 

To achieve the first task, we employ Eq.~(\ref{eq:1}) for the calculation of the function $\langle N \rangle(L)$ at each point of a $401 \times 401$ grid in the $(\ln \zeta,t)$ plane. The grid covers the region $\ln \zeta \in [-0.12,0.12]$ and $t \in [-0.2,0.2]$. The variable $L$, being the ratio of the volume of the entire system (fireball) to the volume of a single nucleon, varies in the range $30 \leq L \leq 700$, corresponding to nuclei with linear size from $3$ to $9$ fm. This range of $L$ values covers all experimentally accessible cases from small to medium size nuclei in \NAFortyNine and \NASixtyOne experiments. 

We perform a linear fit in the function $\ln \langle N \rangle(\ln L)$ at each point of the $(\ln \zeta,t)$ grid and we check its validity with the regression coefficient $R^2$ setting the constraint $R^2 \geq 70\%$ for the characterization as a power-law. The slope parameter of this fit provides us the $\phi_2$-value at each grid point. The set of $(\ln \zeta,t)$ points leading to the same $\phi_2$ value, within a $\pm 10^{-3}$-deviation, determines the curve with constant $\phi_2$ in the $(\ln \zeta,t)$ plane.

\begin{figure}[h]
\includegraphics[width=1.2\textwidth]{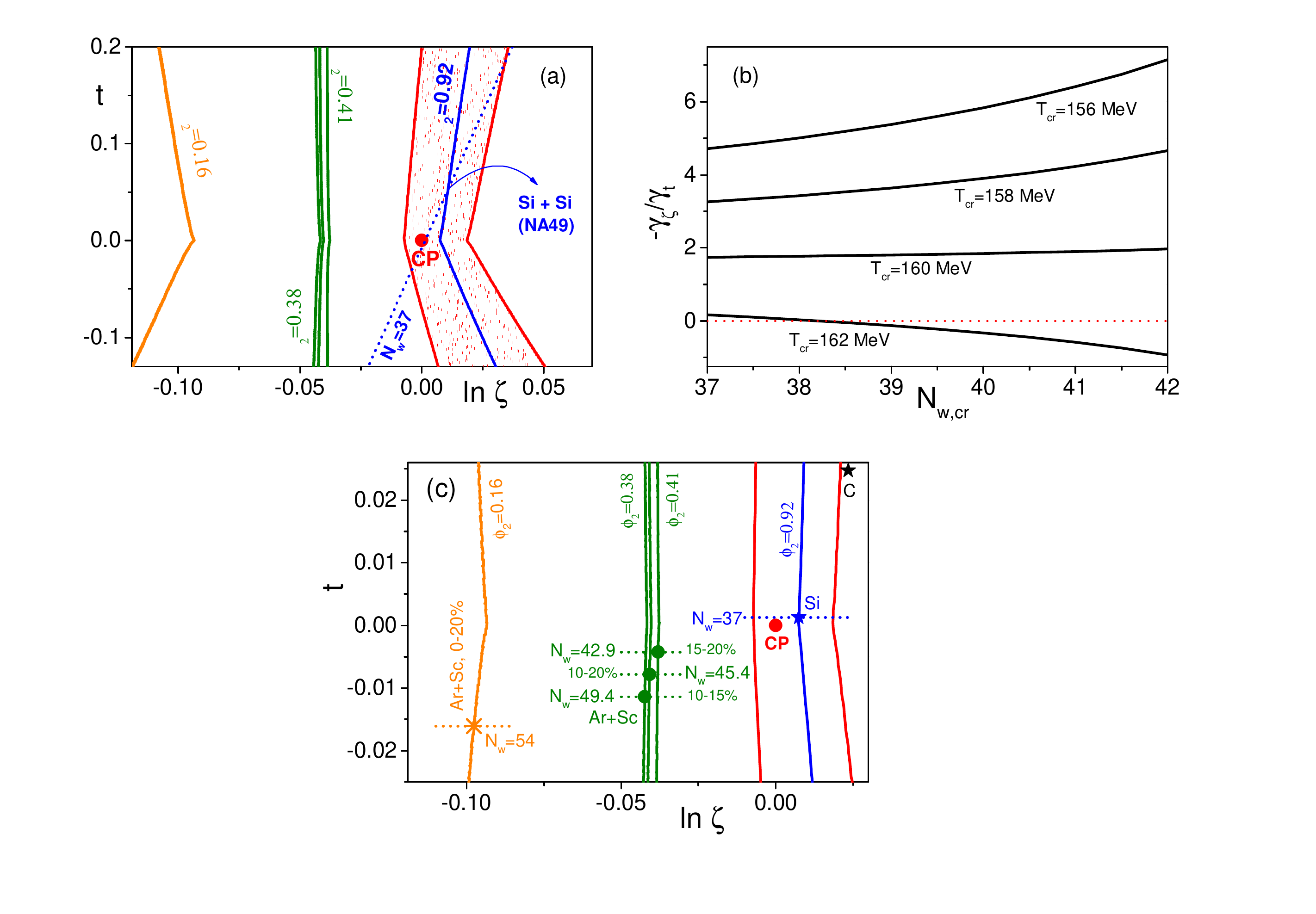}
\caption{(a) The lines $\phi_2=\mathrm{constant}$ in the $(\ln \zeta,t)$ plane. The red lines are the borders of the critical region (red shaded area). The blue line corresponds to $\phi_2=0.92$ associated with the \SiSi freeze-out state (\crange{0}{12} centrality, $158$A~GeV/$c$ collision energy) measured in \NAFortyNine experiment (SPS,CERN). The three dark green lines correspond to $\phi_2=0.41$, $0.39$ and $0.38$ (from right to left) associated with the \ArSc freeze-out states (peripheralities \crange{15}{20},  \crange{10}{20} and \crange{10}{15} respectively) at $150$A~GeV/$c$ (\NASixtyOne experiment). Finally, the orange line corresponds to $\phi_2=0.16$ associated with \ArSc freeze-out state in \crange{0}{20} peripherality. The blue dotted line corresponds to $N_w=37$ line choosing $T_{cr}=156$~MeV and $N_{w,cr}=39.5$. The intersection of the two blue lines defines the location of the \SiSi freeze-out state in the $(\ln \zeta, t)$ plane.
(b) The slope parameter $-\frac{\gamma_{\zeta}}{\gamma_t}$ of the  $N_w=\mathrm{constant}$ lines versus $N_{w,cr}$ for different values of $T_{cr}$.
(c) The lines $\phi_2=\mathrm{constant}$ (coloured, solid) and $N_w=\mathrm{constant}$ (coloured, dotted), for the various systems considered, when $\gamma_{\zeta}=0$ and $T_{cr}=162$~MeV. Their intersection defines the associated freeze-out states. The black star represents the \CC freeze-out state (\NAFortyNine experiment) in the $(\ln \zeta,t)$ plane.}
\label{fig:fig_crit_region_abc}
\end{figure}

 The results of this calculation are shown in Fig.~\ref{fig:fig_crit_region_abc}a. The red lines in this plot correspond to $\phi_2=0.75$ (left) and $\phi_2=1$ (right) determining the borders of the critical region (red shaded area), according to the definition given in \cite{Antoniou_prd_2018}. The blue curve is associated with the $\phi_2$ value for the \SiSi system ($\phi_{2,Si}=0.92$) while the green curves  correspond to the $\phi_2$ values found for \ArSc in the peripherality zones \crange{10}{20}, \crange{10}{15} and \crange{15}{20}, while the orange curve displays the $\phi_2=0.16$ curve (\ArSc, peripherality \crange{0}{20}). For all cases, a piecewise linear description of the $\phi_2(\ln \zeta,t)=\mathrm{constant}$ curve is a good approximation. Notice, that all curves of constant $\phi_2$-values in Fig.~\ref{fig:fig_crit_region_abc}a, being expressed in reduced variables, do not depend on the specific $(\mu_{B,cr},T_{cr})$-values. Thus, the displayed pattern is universal, characterizing the 3d Ising class.

In particular, the information contained in the blue curve in Fig.~\ref{fig:fig_crit_region_abc}a can be further exploited, when combined with the results plotted in Fig.~\ref{fig:phi2_vs_Nwounded}, as well as the data for the freeze-out baryochemical potential $\mu_{B,Si}$ and temperature $T_{Si}$ of the \SiSi system, estimated with the statistical hadronization model in \cite{Becattini_2006}. In fact, it allows the determination of the critical baryochemical potential $\mu_{B,cr}$ as a function of the critical temperature $T_{cr}$, following the line of thought given below:

\begin{itemize}

\item Fig.~\ref{fig:phi2_vs_Nwounded} dictates that the critical temperature is slightly below the freeze-out temperature of the \SiSi system, since the function $T(N_w)$ is decreasing \cite{Becattini_2006,Becattini_2014}. This means that the freeze-out state of \SiSi must lie on the upper branch of the blue curve in Fig.~\ref{fig:fig_crit_region_abc}a where $t_{Si}=\frac{T_{Si}-T_{cr}}{T_{cr}} > 0$.

\item The upper branch of the blue curve in Fig.~\ref{fig:fig_crit_region_abc}a can be described by the line:
\begin{equation}
t=a_{Si} \ln \zeta + b_{Si}~~~~;~~~~a_{Si}=47.52~;~b_{Si}=-0.36
\label{eq:3}
\end{equation}
obtained by fitting.

\item For given $\mu_{B,Si}$ and $T_{Si}$, equation (\ref{eq:3}) provides a relation between $\mu_{B,cr}$ and $T_{cr}$:
\begin{equation}
\mu_{B,cr}=\mu_{B,Si}+ \displaystyle{\frac{(1+b_{Si}) T_{cr}-T_{Si}}{a_{Si}}}
\label{eq:4}
\end{equation}
taking into account that $\ln \zeta_{Si} = \frac{\mu_{B,Si}-\mu_{B,cr}}{k_B T_{cr}}$.

\item In \cite{Becattini_2006} it is found  that $\mu_{B,Si}=260\pm 17.9$~MeV and $T_{Si}=162.2 \pm 7.9$~MeV. Using the central values $(\mu_{B,Si},T_{Si})=(260,162.2)$ MeV and assuming $T_{cr}=162$ MeV, we find $\mu_{B,cr}=258.8$ MeV. Although this is an indicative result obtained neglecting errors, the described procedure can be used for any set of $\{\mu_{B,Si},T_{Si},T_{cr}\}$ values to determine $\mu_{B,cr}$. Notice that the choice $T_{cr}=162$ MeV is compatible with some recent Lattice QCD results \cite{Datta_Lattice}.

\end{itemize}

In the second step we place the curves $N_w(\ln \zeta,t)=c_i$ ($i=1,..,5$) on the $(\ln \zeta,t)$-plane. We assume, therefore, a lowest order expansion of the function $N_w(\ln \zeta,t)$ around $(\ln \zeta_{Si}, t_{Si})$ of the form:
\begin{equation}
N_w(\ln \zeta,t)=N_w(\ln \zeta_{Si}, t_{Si}) + \gamma_t (t-t_{Si}) + \gamma_{\zeta} (\ln \zeta - \ln \zeta_{Si})
\label{eq:5}
\end{equation}
with $N_w(\ln \zeta_{Si}, t_{Si})=37$ (see Table~\ref{tab:AMIAS_Nw_phi2}) and $\gamma_t$, $\gamma_{\zeta}$ constants to be determined. Notice that, according to Fig.~\ref{fig:phi2_vs_Nwounded} (and Table~\ref{tab:AMIAS_Nw_phi2}), the number of wounded nucleons $N_{w,cr}$ at the critical point $(\ln \zeta_{cr},t_{cr})=(0,0)$ is bounded in the range $37 \leq N_{w,cr} < 42.4$. Therefore it is convenient, if possible, to parametrize the constants $\gamma_t$ and $\gamma_{\zeta}$ in Eq.~(\ref{eq:5}) in terms of $N_{w,cr}$. To this end, we use the  freeze-out state $(\ln \zeta_C, t_C)$ formed in central \CC collisions at 158A~GeV/$c$ (\NAFortyNine, CERN), which lies close enough to the \SiSi freeze-out state \cite{Becattini_2006} to obey Eq.~(\ref{eq:5}). We obtain the system of equations:
\begin{subequations}
\begin{align}
&N_{w,cr}=N_{w,Si} - \gamma_t t_{Si} - \gamma_{\zeta} \ln \zeta_{Si}\label{eq:6a}\\
&N_{w,C}=N_{w,Si} + \gamma_t (t_{C}-t_{Si}) + \gamma_{\zeta} (\ln \zeta_{C} - \ln \zeta_{Si}) \label{eq:6b}
\end{align}
\end{subequations}
with $N_{w,Si}=N_w(\ln \zeta_{Si}, t_{Si})$ and $N_{w,C}=N_w(\ln \zeta_{C}, t_{C})$ respectively. Eqs.~(\ref{eq:6a},\ref{eq:6b}) can be solved for $\gamma_t$ and $\gamma_{\zeta}$ yielding:
\begin{subequations}
\begin{align}
&\gamma_t=\frac{N_{w,Si} - N_{w,cr} - \gamma_{\zeta}\ln \zeta_{Si}}{t_{Si}} \label{eq:7a}\\
&\gamma_{\zeta}=\frac{t_{Si} (N_{w,C} - N_{w,Si}) + (t_C - t_{Si}) (N_{w,cr}-N_{w,Si})}
{t_{Si} (\ln \zeta_C - \ln \zeta_{Si}) - \ln \zeta_{Si} (t_C - t_{Si})} \label{eq:7b}
\end{align}
\end{subequations}
In Eqs.~(\ref{eq:7a},\ref{eq:7b}) $N_{w,Si}$ and $N_{w,C}$ are given in \cite{NA49_wounded} while $t_{Si}$, $\ln \zeta_{Si}$, $t_C$ and $\ln \zeta_C$ are determined by the values of $T_{Si}$, $\mu_{B,Si}$, $T_C$ and $\mu_{B,C}$ given in \cite{Becattini_2006} provided that $T_{cr}$ and $\mu_{B,cr}$ are known. In consistency with the analysis in the first step we will use only the central values of \cite{Becattini_2006} for $T_{Si}$, $\mu_{B,Si}$, $T_C$ and $\mu_{B,C}$ (neglecting errors).

Furthermore, $\mu_{B,cr}$ can be expressed in terms of $T_{cr}$ via Eq.~(\ref{eq:4}). Thus, the only unknown parameters are $T_{cr}$ and $N_{w,cr}$. Their values determine the lines $N_w=\mathrm{constant}$. In Fig.~\ref{fig:fig_crit_region_abc}a we show as an example the line $N_w=37$ (dotted blue) assuming $T_{cr}=156$ MeV and $N_{w,cr}=39.5$. According to Table~\ref{tab:AMIAS_Nw_phi2} this line corresponds to the \SiSi system. The intersection of this line with the curve $\phi_2=0.92$ determines the location of the \SiSi freeze-out state in the $(\ln \zeta,t)$ plane. The blue arrow is used to clarify this. In a similar way we can find the locations of the freeze-out states for the \ArSc systems at different peripheralities. They depend strongly on the chosen values $T_{cr}$ and $N_{w,cr}$. Varying $T_{cr}$ and $N_{w,cr}$ the slope $-\frac{\gamma_{\zeta}}{\gamma_t}$ of the lines $N_w=\mathrm{constant}$ changes. A more quantitative description of this dependence is displayed in Fig.~\ref{fig:fig_crit_region_abc}b where we plot $-\frac{\gamma_{\zeta}}{\gamma_t}$ as a function of $N_{w,cr}$ for different values of $T_{cr}$. We observe that this dependence becomes smoother, in the sense that the slope $-\frac{\gamma_{\zeta}}{\gamma_t}$ becomes smaller, for $T_{cr}=162$ MeV. In the following considerations we will exclusively use this value for the critical temperature. With this choice, the data used to calculate the constants $\gamma_t$ and $\gamma_{\zeta}$ get the values summarized in Table~\ref{tab:AA_T_muB}.

\begin{table}[h]
\begin{center}
\begin{tabular}{| l | c | c | c | c | c |}
\hline
\AnyReaction & $N_w$ & $(\mu_B,T)$ (MeV) & $(\ln \zeta,t)$ & $\mu_{B,cr}$ (MeV) &
$T_{cr}$ (MeV)\\
\hline
\CC & 14 & (262.6,166) & (0.0235,0.0247) &  &  \\
\SiSi & 37 & (260,162.2) & (0.007,0.001) & 258.8 & 162 \\
\hline
\end{tabular}
\end{center}
\caption{Calculation of constants $\gamma_t$, $\gamma_{\zeta}$ for the choice of $T_{cr}=162$~MeV.}
\label{tab:AA_T_muB}
\end{table}

The entries of Table~\ref{tab:AA_T_muB} determine the constants $\gamma_t$, $\gamma_{\zeta}$ in terms of $N_{w,cr}$ as:
\begin{equation}
\gamma_t= -4833.9 + 100.8 N_{w,cr}~~~~;~~~~\gamma_{\zeta}= 5631.9 - 147.4 N_{w,cr}
\label{eq:8}
\end{equation}
Then, it is straightforward to obtain the lines of constant value of $N_w$ in the $(\ln \zeta,t)$ plane using Eq.~(\ref{eq:5}), after choosing $N_{w,cr} \in [37,42.4)$.

Especially interesting, from the phenomenological point of view, is the case where $N_w$ is independent of $\ln \zeta$. Such a behaviour is also suggested by experimental data \cite{Becattini_2006,Becattini_2014}. Within our analysis, this scenario implies that $\gamma_{\zeta}$ in Eq.~(\ref{eq:7b}) vanishes, leading to the condition:
\begin{equation}
N_{w,cr}=N_{w,Si} + \frac{(T_{Si} - T_{cr}) (N_{w,Si}-N_{w,C})}{T_{C}-T_{Si}}
\label{eq:9}
\end{equation}
which connects the critical temperature $T_{cr}$ with the number of wounded nucleons at the critical point $N_{w,cr}$. For  $T_{cr}=162$ MeV we find, using Eq.~(\ref{eq:9}), $N_{w,cr}=38.2$. Using this value for $N_{w,cr}$ we show in Fig.~\ref{fig:fig_crit_region_abc}c the critical region and its neighbourhood in the $(\ln \zeta, \phi_2)$-plane. The location of the \ArSc freeze-out states for different peripheralities (\NASixtyOne experiment) in the interval \crange{10}{20} is shown by the green circles in this plot. The blue star displays the location of the \SiSi freeze-out state, while, for completeness, we include in the graph also the \CC freeze-out state (black star), both measured in \NAFortyNine experiment. The red circle indicates the location of the critical point. Finally the orange cross presents the freeze-out state of \ArSc in \crange{0}{20} peripherality interval. It is remarkable that the freeze-out states of the different systems are concentrated close to the critical point along the reduced temperature axis. At this scale the $\phi_2=\mathrm{constant}$ curves appear as almost vertical lines forming, together with the horizontal $N_w=\mathrm{constant}$ (dotted coloured) lines a cartesian coordinate system for the description of the critical region and its neighbourhood. 

Within this parametrization a nice interpretation is possible: the number of wounded nucleons $N_w$ defines the temperature of a freeze-out state within and close to the critical region while the intermittency index $\phi_2$ defines the corresponding baryochemical potential. As already discussed, small changes in the $T_{cr}$ and in the $T_{Si}$, $\mu_{B,Si}$, $T_C$, $\mu_{B,C}$ values will not alter the form of the critical neighbourhood shown in Fig.~\ref{fig:fig_crit_region_abc}c since the plot is in relative (reduced) coordinates with respect to the critical point. Thus, similarly to Fig.~\ref{fig:fig_crit_region_abc}a, also Fig.~\ref{fig:fig_crit_region_abc}c is a rather universal result. It indicates clearly the proximity to the critical region of the peripheral \ArSc collisions. However, within the peripherality range \crange{0}{20}, allowed by the \NASixtyOne experiment, the recorded freeze-out states are still at a distance from the critical region and therefore the observed intermittency effect \cite{Davis_PPC_2019} can not be characterized as critical. Nevertheless, together with \NAFortyNine intermittency results \cite{NA49_pions,NA49_protons}, the \NASixtyOne intermittency analysis provides a close frame of the critical region.

\section{Summary and conclusions}
\label{sec:sec5}

Employing the AMIAS method to extract the distributions of the intermittency index $\phi_2$ from the preliminary \NASixtyOne results for the proton correlator $\Delta F_2(M)$ in transverse momentum space of the \ArSc system ($150$A~GeV/$c$ beam momentum) at different peripheralities \cite{Davis_PPC_2019}, we demonstrate the presence of a non-vanishing intermittency effect in the \crange{10}{20} peripherality interval. The AMIAS method of analysis is also applied to the $\Delta F_2(M)$ measurement in \SiSi central collisions (\crange{0}{12} peripherality) at $158$A~GeV/$c$ by the \NAFortyNine experiment \cite{NA49_protons}, verifying the presence of critical fluctuations in this system. The AMIAS results for $\phi_2$, combined with an estimation of the corresponding number of wounded nucleons $N_w$ for each considered freeze-out state, allow the mapping of the critical region and its neighbourhood in the reduced baryochemical-temperature plane, in terms of the quantities $\phi_2$ and $N_w$. This mapping indicates that the preliminary \NASixtyOne intermittency results are fully compatible with the corresponding \NAFortyNine measurements reflecting transparently the approach of the \ArSc freeze-out states towards the critical region with increasing peripherality. Our analysis provides strong constraints on the location of the critical region and it may be used as an invaluable guide in the forthcoming experimental searches for the QCD critical point.

{\small \textbf{Acknowledgments:}  This work was supported by the National Science Centre, Poland (grant no. 2014/14/E/ST2/00018).}

\end{document}